\documentclass[aps,prd,fleqn,superscriptaddress]{revtex4}
\usepackage{graphicx,xcolor,natbib,braket,float}
\usepackage{amsmath,amssymb,amsfonts}
\newcommand{\bse}{\begin{subequations}}
\newcommand{\ese}{\end{subequations}}
\newcommand{\be}{\begin{equation}}
\newcommand{\ee}{\end{equation}}
\newcommand{\bea}{\begin{eqnarray}}
\newcommand{\eea}{\end{eqnarray}}
\newcommand{\ba}{\begin{array}}
\newcommand{\ea}{\end{array}}

\usepackage[colorlinks=true, linkcolor=blue, bookmarks=true]{hyperref}
\begin{document}
\title{Velocity dependence of holographic entanglement entropy in a charged plasma}
\author{V. Esrafilian\footnote{$\rm{v}_{-}$esrafilian@sbu.ac.ir}}
\author{M. Ali-Akbari\footnote{$\rm{m}_{-}$aliakbari@sbu.ac.ir}}
\affiliation{Department of Physics, Shahid Beheshti University, 1983969411, Tehran, Iran}

\begin{abstract}
We have studied holographic entanglement entropy in a moving thermal gauge theory with a non-zero chemical potential. A sufficiently high velocity enhances the holographic entanglement entropy, particularly for larger values of the chemical potential. However, at high temperature and velocity, the chemical potential dependence is almost entirely washed out, indicating that thermal fluctuations dominate over charge-density effects. In the ultrarelativistic regime, the holographic entanglement entropy grows very rapidly with velocity, which emerges as the dominant parameter, largely suppressing both thermal and chemical contributions.\end{abstract}

\maketitle
\tableofcontents

\section{Introduction}
Gauge-gravity duality has, over the last two decades, become an instructive and applicable tool for tackling strongly coupled gauge theories. This duality is a strong-weak duality, stating that a strongly coupled gauge theory can be described by a classical gravity theory in a higher dimension, where the gauge theory lives on the boundary of the gravity side \cite{Maldacena:1997re,Witten:1998qj,Gubser:1998bc,Aharony:1999ti}. Consequently, many important quantities and processes in the gauge theory have a counterpart in the gravity side. For this reason, the duality has been widely applied to compute or explain various complicated physical quantities and processes, such as phase transitions, out-of-equilibrium processes and thermalization \cite{Fischler:2012uv,Ali-Akbari:2013hba,Witten:1998zw,Ahmed:2023dnh,Matsumoto:2018ukk,Vahedi:2018gvn,Lindgren:2017hiu}. In fact, efforts have been made to use the duality to provide an acceptable understanding of these physical quantities and to explain how to compute or describe their evolution. This has led to the study of very different areas of physics. One of the quantities in information theory that has attracted a great deal of attention is entanglement entropy (EE) \cite{Headrick:2019eth,Sorkin:1984kjy,Calabrese:2004eu,Casini:2009sr}.

EE quantifies how two quantum systems are entangled with each other, or equivalently, how much information is lost when the degrees of freedom of one subsystem are traced out. To describe it mathematically, consider a Hilbert space partitioned into two parts, $A$ and its complement $A^c$, such that $\mathcal{H} = \mathcal{H}_A \otimes \mathcal{H}_{A^c}$. Suppose a pure state $\ket{\psi}$ describes the physical system with density matrix $\rho = \ket{\psi}\bra{\psi}$. By tracing over the degrees of freedom of the complementary subsystem, we obtain the reduced density matrix $\rho_A = \operatorname{Tr}_{A^c}(\rho)$. The entanglement entropy is then defined as $S = -\operatorname{Tr}(\rho_A \log \rho_A)$. A vanishing value of this quantity indicates that the two subsystems are disentangled while any non-zero value quantifies the degree of entanglement between them. EE satisfies several important properties, such as subadditivity and strong subadditivity, and for a pure state it is symmetric: $S(\rho_A)=S(\rho_{A^c}) $ \cite{Headrick:2007km,Headrick:2013zda}.

\begin{figure}[H]
  \centering
  \includegraphics[width=0.6\textwidth]{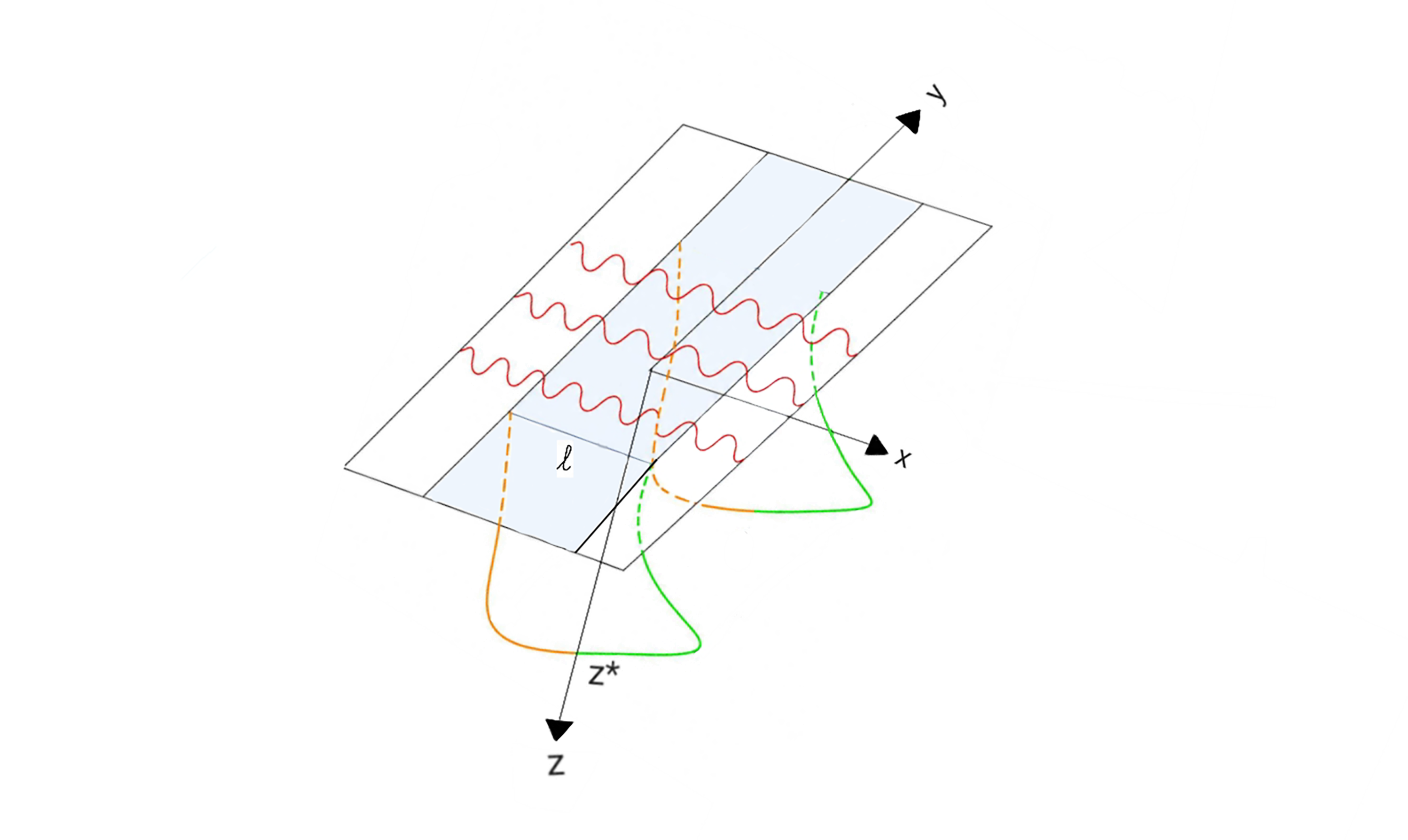}
   \caption{Schematic illustration of the subsystem $A$ and its complement $A^c$.}
\label{general}
\end{figure}

Although EE is difficult to compute, especially in strongly coupled gauge theories, gauge-gravity duality provides a remarkably simple proposal to study it, first introduced by Ryu and Takayanagi (RT) \cite{Ryu:2006bv}. Their proposal states that the EE is given by the area of a minimal (or more generally extremal) surface extended in the bulk, homologous to \(A\) and anchored to the entangling surface between the subsystem \(A\) and its complement on the boundary of the gravity theory, see figure \ref{general}. The proposal has passed various checks and is widely accepted to correctly describe EE. Consequently, using this proposal, various aspects of EE in strongly coupled gauge theories have been studied and reported in the literature. Through this holographic dictionary, EE has been successfully employed to probe a variety of phenomena, including quantum phase transitions, entanglement dynamics in out-of-equilibrium processes, and other non-equilibrium phenomena \cite{Headrick:2010zt,Tonni:2010pv,Fischler:2012uv,Molina-Vilaplana:2011ydi,Fischler:2013gsa,Ling:2015dma}.

The combined effect of a velocity and a chemical potential on the holographic entanglement entropy (HEE) has received less attention which is the main focus of the present work. Here, we systematically study the HEE in a boosted, charged black hole background and show how velocity and chemical potential jointly affect the entanglement structure. Our numerical results demonstrate that both parameters enhance the HEE, albeit in in a way that compensates each other. 

\section{Review on the background}
The background we consider in this paper is the Reissner--Nordstr\"om--AdS black hole whose four-dimensional metric is given by
\begin{equation}\label{metric}
ds^2 = \frac{1}{z^2}\left(-f(z)\,dt^2 + \frac{dz^2}{f(z)} + dx^2 + dy^2\right),
\end{equation}
where
\begin{equation}
\begin{split}
f(z) &= 1 - M z^3 + \frac{Q^2}{2}\,z^4,\\
A_t(z) &= Q\,(z_h - z),
\end{split}
\end{equation}
where the AdS radius is set to be one. This metric is a solution of four-dimensional Einstein-Maxwell theory with a negative cosmological constant \cite{Kundu:2016dyk}. The coordinate $z$ is the radial direction and according to gauge-gravity duality the $(2+1)$-dimensional gauge theory lives on the boundary of this background at $z\to 0$. The black hole horizon is located at $z = z_h$ which satisfies $f(z_h)=0$. Consequently, the temperature of the black hole, which is identified with the temperature of the gauge theory, is given by
\begin{equation}
T = \frac{3}{4\pi z_h}(1-\frac{Q^2 z_h^4}{6}).
\end{equation}
The parameters $M$ and $Q$ describe the mass and the electric charge of the black hole, respectively. Since the black hole is charged, the field theory living on the boundary acquires a non-zero chemical potential
\begin{equation}
\mu = Q z_h.
\end{equation}
This family of solutions has been extensively studied in the literature; see, for example, \cite{Chamblin:1999tk,Cvetic:1999ne,Cai:1996eg}. It is clear that the above metric reduces to the standard four-dimensional AdS black hole background when $Q = 0$.

What we would like to examine here is the effect of velocity on the HEE. Thus, we apply a boost in the $x$-$t$ directions to the metric \eqref{metric} and finally obtain
\be %
ds^2=\frac{1}{z^2}\left(-dt^2+g(v,z)(dt-vdx)^2+\frac{dz^2}{f(z)} + dx^2+ dy^2\right),
\ee
where
\be
g(v,z)=\gamma^2\left(1-f(z)\right),
\ee 
and $\gamma=(1-v^2)^{-1/2}$. On the gauge theory side, the velocity $v$ obviously describes a moving plasma.

Based on the RT proposal \cite{Ryu:2006bv}, to compute the HEE, we consider a strip-like subsystem embedded in the three-dimensional spacetime at a fixed value of time, say $t=0$. In the presence of a velocity, the extremal surface is chosen as (see figure \ref{general})
\begin{equation}
x = x(z),\quad t = t(z),\quad -\frac{L}{2} \leq y \leq \frac{L}{2},
\end{equation}
where $L \rightarrow \infty$ denotes the infinite transverse direction and the short edge of the strip, $l$, is aligned with the direction of the boost, i.e., the $x$-direction. Note that for sufficiently large $L$, translational symmetry along the $y$-direction is preserved. Therefore, we need to extremize the following area functional
\begin{equation}
S = L \int \frac{dz}{z^2} \sqrt{-(1-g)\,t^{\prime 2} + (1+v^2 g)\,x^{\prime 2} - 2v g\,t^\prime x^\prime + f^{-1}},
\end{equation}
where the prime denotes differentiation with respect to $z$. Treating the integrand as a Lagrangian, one can easily find the equations of motion for $x(z)$ and $t(z)$. Since the equations of motion are lengthy, we do not present them here explicitly. To solve these equations we consider two branches of solutions with the following boundary conditions:
\begin{equation}
x_\pm(z \rightarrow 0) = \pm \frac{l}{2},\quad t_\pm(z \rightarrow 0) = 0,\quad x^\prime_\pm(z \rightarrow z_*) = \infty,\quad t^\prime_\pm(z \rightarrow z_*) = \infty.
\end{equation}
Here $z_*$ denotes the turning point of the extremal surface, as illustrated in figure \ref{general}. In the next section, we solve the equations of motion subject to the boundary conditions numerically, compute the value of the HEE in the presence of velocity and discuss its dependence on the parameters of the problem at hand, namely the chemical potential and the temperature.

\section{Numerical results}
To present our numerical results, we consider three distinct classes of solutions, characterized by the chemical potential, temperature and velocity. The various dependencies of the HEE on these parameters will be investigated. In the following results, we set $l=1$. We have also examined other values of $l$ and, up to our numerical precision, we observe the same qualitative behavior.

\subsection{Velocity (Chemical potential) dependence at fixed chemical potential (velocity) and temperature}
\begin{figure}[H]
  \centering
  \includegraphics[width=0.49\textwidth]{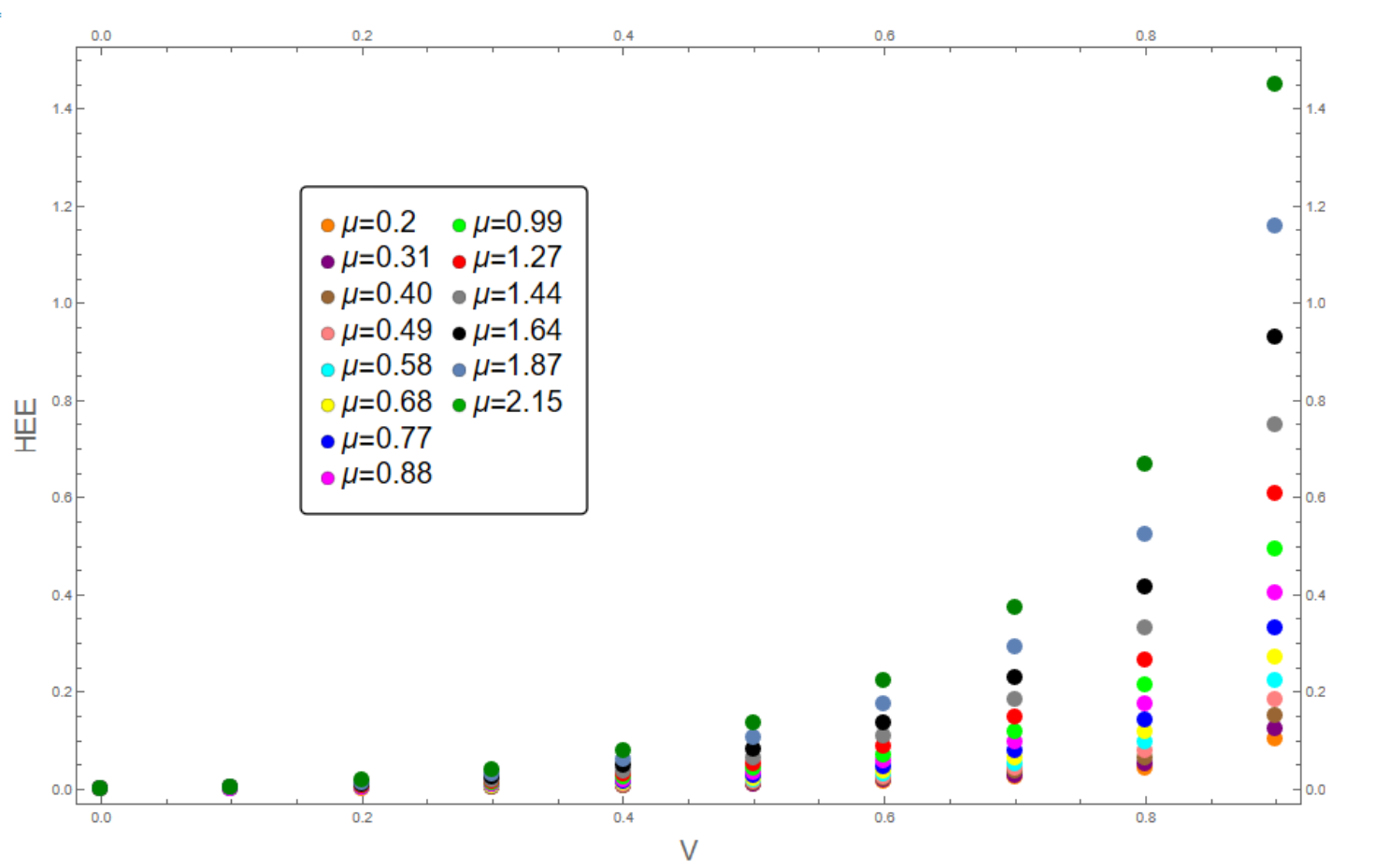}
  \includegraphics[width=0.49\textwidth]{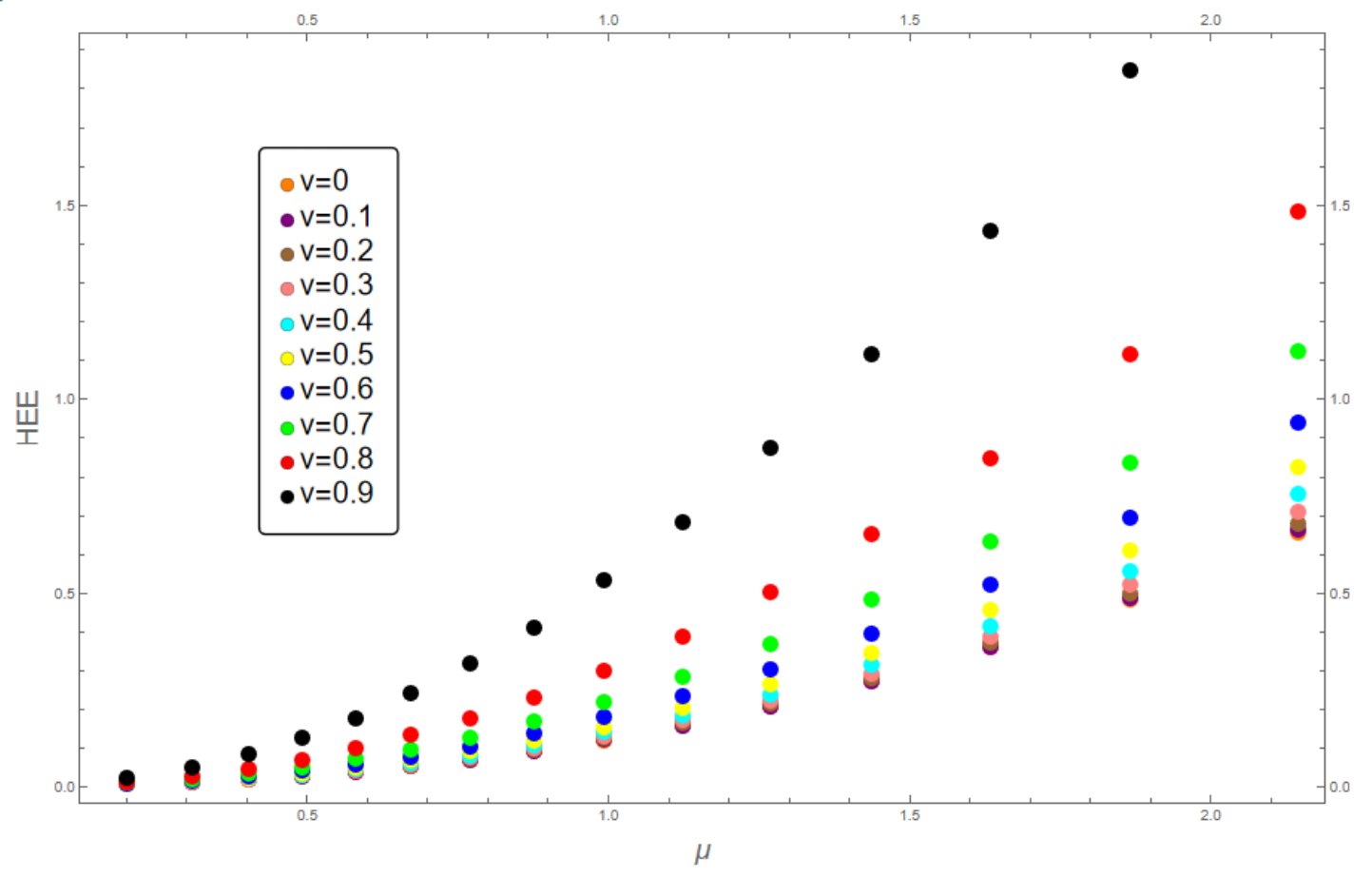}
  \caption{Left: The HEE versus velocity for fixed chemical potential values.\  Right: The HEE versus chemical potential for fixed velocity values. The temperature is set to $T=0.1$.}\label{figvmu0}
\end{figure}

In figure \ref{figvmu0}, left, the HEE is plotted as a function of velocity. To obtain this figure, we define the quantity
\begin{equation}\label{sub1}
E_v \equiv \frac{1}{L}\left[S(\mu,T,v) - S(\mu,T,v=0)\right],
\end{equation}
which guarantees that the effect of velocity is isolated for given values of chemical potential and temperature.

Evidently, the HEE increases with velocity as first studied in \cite{Bhattacharya:2022msw}. The figure also shows that for larger values of the chemical potential, this increase becomes more substantial. In other words, to observe the effect of velocity on the HEE, a sufficiently large chemical potential is required, $\mu\gtrsim 0.5$. Moreover, for a given value of the HEE, velocity and chemical potential have opposite effects: a larger velocity requires a lower chemical potential to produce the same HEE.

Our next step is to study the effect of the chemical potential on the HEE. To this end, we define the quantity
\begin{equation}\label{sub2}
E_\mu \equiv \frac{1}{L}\left[S(\mu,T,v) - S(\mu=0,T,v)\right],
\end{equation}
which isolates the effect of the chemical potential for given values of velocity and temperature. 
Our numerical results are shown in figure \ref{figvmu0}, right. Similar to the previous subsection, the HEE increases as the chemical potential increases. Furthermore, the effect of the chemical potential is very substantial at high velocities. However, for small values of the chemical potential, $\mu\lesssim 0.5$, the effect remains insignificant even at high velocity. Again, the effects of velocity and chemical potential compensate each other for a given value of the HEE.


At fixed temperature and for given values of chemical potential and velocity, let us say $\mu_0$ and $v_0$, the final result for the HEE obtained from \eqref{sub1} and \eqref{sub2} must be the same. Actually, we describe a unique physical system using two different parameters. But the point we should notice is that the subtraction terms in \eqref{sub1} and \eqref{sub2} are different. Therefore, as a cross-check, we compare our numerical results and observe
\[
E_\mu(\mu_0,v_0) + \frac{1}{L} S(\mu=0,T,v_0) = E_v(\mu_0,v_0) + \frac{1}{L} S(\mu_0,T,v=0).
\]
This equality must hold for any consistent numerical implementation. The left-hand side uses the chemical-potential subtraction scheme while the right-hand side employs the velocity-subtraction scheme. If the equality is satisfied within numerical precision, it confirms that our calculations correctly capture the physical HEE independent of the chosen subtraction method. Any significant deviation would indicate an inconsistency in the numerical integration or in the handling of the divergent terms.

\subsection{Temperature dependence at fixed chemical potential and velocity}
\begin{figure}[H]
  \centering
  \includegraphics[width=0.49\textwidth]{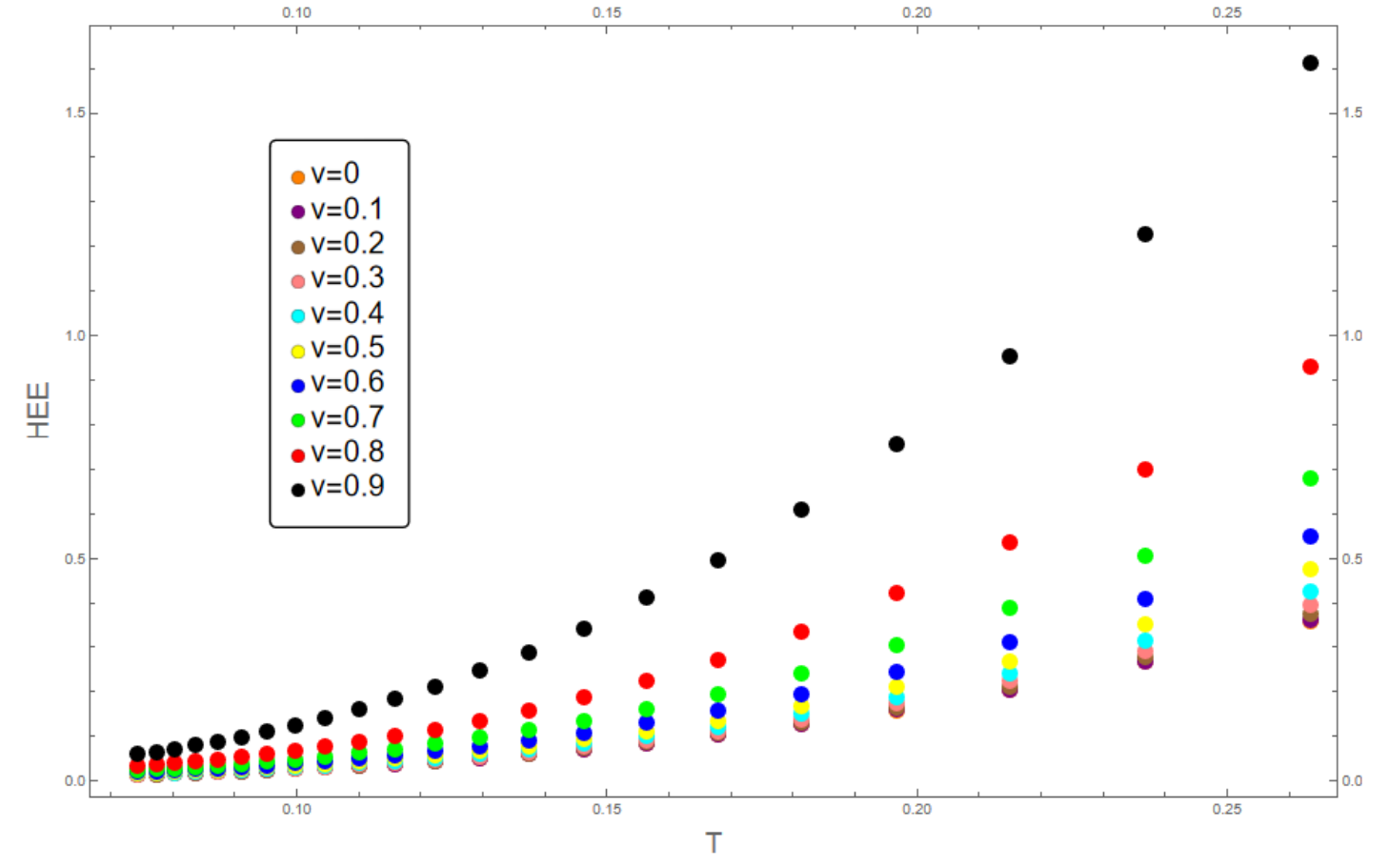}
 \includegraphics[width=0.49\textwidth]{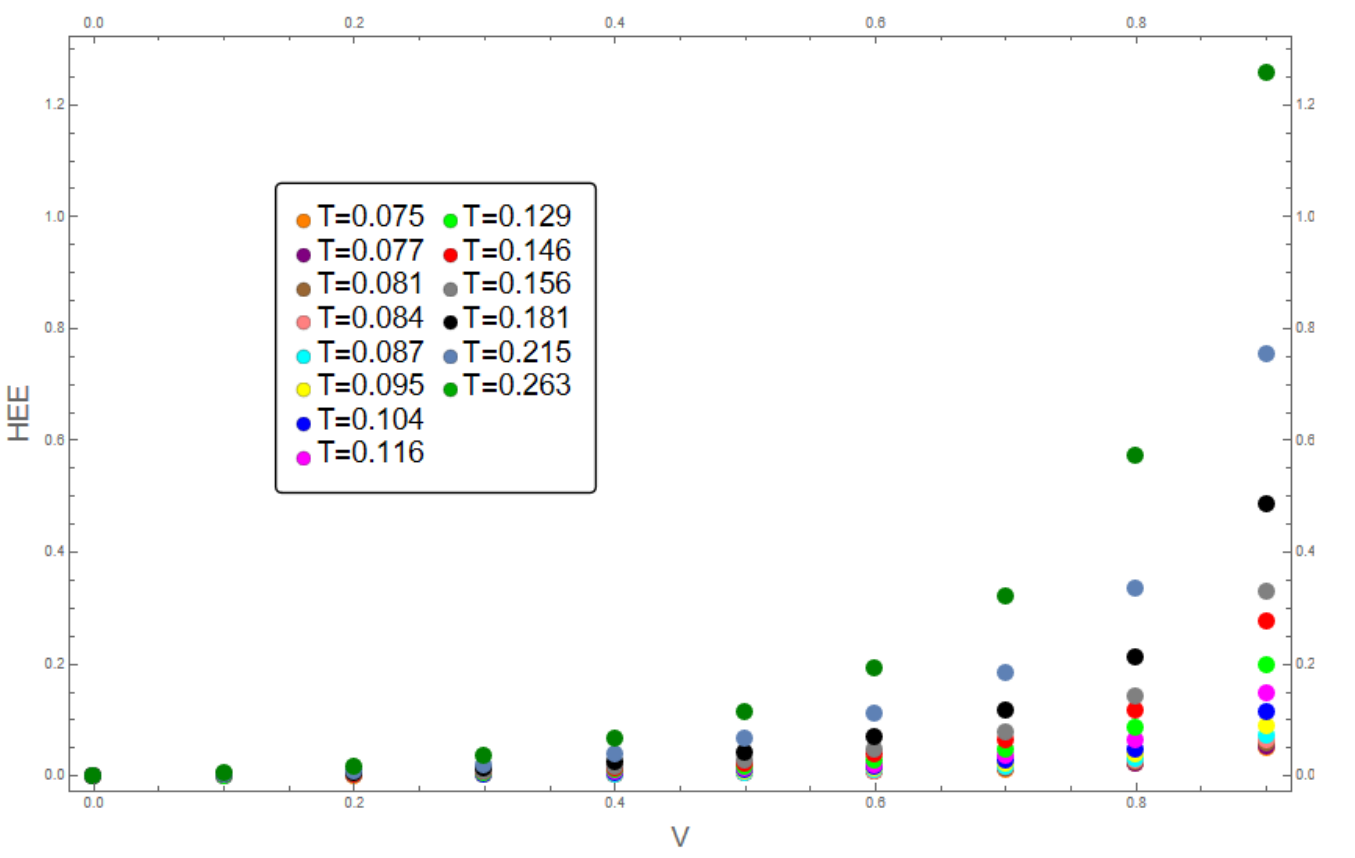}
    \caption{Left: The HEE versus temperature for fixed velocity values.\  Right: The HEE versus velocity for fixed temperature values. The chemical potential is set to $\mu=0.203$.}\label{figvmu}
\label{figtem}
\end{figure}

The temperature dependence of HEE has been extensively studied in the literature for various holographic physical systems, for instance see \cite{Kundu:2016dyk,Saha:2019ado,Ebrahim:2023ush,Rahimi:2018ica,Bah:2007kcs}. Therefore, here we briefly report our results and emphasize the velocity effect. Figure~\ref{figtem} shows that HEE increases by raising the temperature at fixed $v$ or by raising the velocity at fixed temperature. In fact, similar to our previous results, the left and right panels in this figure describe the same physical system. It is clearly seen that for large enough temperature, $T\gtrsim 0.2$, the effect of velocity on HEE is significant. To obtain this figure, we define the quantities
\begin{align}
E^{(1)}_T &= \frac{1}{L}\left[S(\mu,T,v) - S(\mu,T=0,v)\right],
\end{align}
and
\begin{align}
E^{(2)}_T &= \frac{1}{L}\left[S(\mu,T,v) - S(\mu,T,v=0)\right],
\end{align}
for the left and right panels, respectively.
This definition isolates the contribution from either temperature or velocity alone, by subtracting the reference configuration where one of them vanishes. The equality of the two expressions serves as a consistency check: both subtractions should yield the same physical excess HEE. At low temperatures, $T\lesssim 0.1$, the velocity effect is suppressed, whereas at high temperatures, thermal fluctuations dominate and the velocity further enhances the HEE. The monotonic increase observed in figure \ref{figtem} is consistent with the expectation that both thermal and flow excitations add independent entangled modes across the entangling surface.

\begin{figure}[H]
  \centering
  \includegraphics[width=0.5\textwidth]{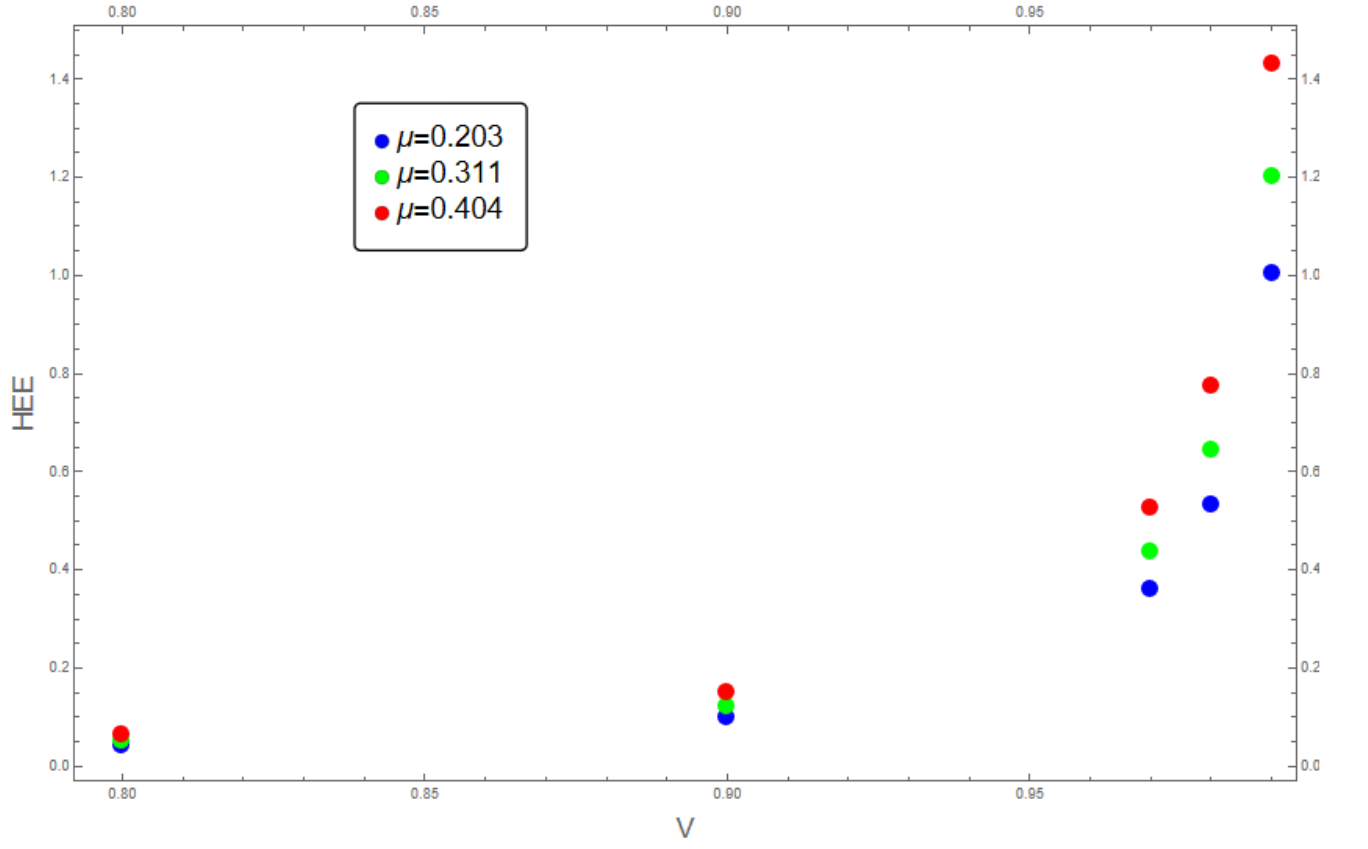}
\caption{HEE as a function of the velocity $v$ in the ultrarelativistic regime for three distinct values of the chemical potential $\mu$ and $T=0.1$.}\label{last}
\end{figure}

In figure \ref{last}, we examine the ultrarelativistic regime, specifically $v > 0.9$, and plot the HEE as a function of velocity for three representative values of the chemical potential. In this limit, the HEE exhibits significant and rapid growth with increasing velocity. Notably, the velocity emerges as the dominant control parameter for the HEE at very high boosts. This behavior indicates that the velocity $v$ strongly enhances the holographic area while the influence of the chemical potential is comparatively less pronounced than the boost effect. As a result, the ultrarelativistic dynamics seem to be primarily governed by the velocity rather than by the charge density or temperature.

\begin{figure}[H]
  \centering
 \includegraphics[width=0.48\textwidth]{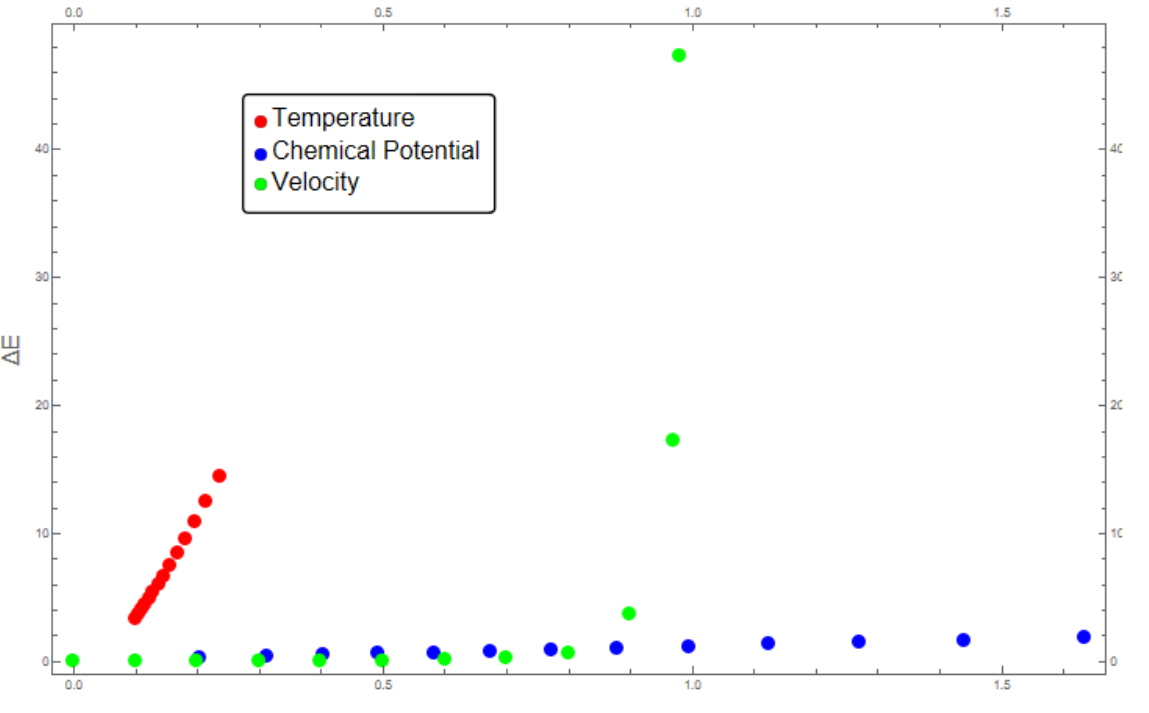}
 \caption{$\Delta E$ as a function of temperature for fixed chemical potential $\mu = 0.203$ and velocity $v=0.9$ (red), as a function of chemical potential for fixed temperature $T = 0.1$ and velocity $v=0.9$ (blue) and as a function of velocity for fixed temperature $T = 0.1$ and chemical potential $\mu=0.203$ (green). Our starting point is identified by $\mu = 0.203$, $T=0.1$ and $v=0.9$ and we then increase the chemical potential, temperature and velocity, separately.
}\label{HEEvar}
\end{figure}

To assess the relative influence of the chemical potential, temperature and velocity on the HEE, we compute the corresponding finite-difference slopes and display them in the figure \ref{HEEvar}. The slopes with respect to the chemical potential, temperature and velocity are defined as
\bse\begin{align}
\Delta E_\mu &= \frac{E(\mu_f) - E(\mu_i)}{\mu_f - \mu_i},\\
\Delta E_T &= \frac{E(T_f) - E(T_i)}{T_f - T_i},\\
\Delta E_v &= \frac{E(v_f) - E(v_i)}{v_f - v_i},
\end{align}\ese
respectively. Our numerical results clearly indicate that, for $v \lesssim 0.9$, the HEE is significantly more responsive to temperature variations than to changes in the chemical potential or velocity. In particular, at sufficiently high temperatures, the dependence on the chemical potential becomes essentially negligible. In the ultrarelativistic regime, however, the velocity effect becomes more significant than both temperature and chemical potential variations, ultimately governing the behavior of the HEE.


\section{Remarks}

To explain the increase of HEE with velocity, recall that the boost direction is chosen along the short edge of the strip $l$, i.e. perpendicular to the entangling surface. Consider the Lorentz transformation from lab coordinates $(t,x)$ to the plasma rest frame coordinates $(t',x')$:
\be
t' = \gamma(t - v x), \qquad x' = \gamma(x - v t).
\ee
When $v = 0$, we have $t' = t$ and $x' = x$: the equal-time slice in the lab (say $t = 0$) coincides with the equal-time slice $t' = 0$ in the rest frame, a flat, horizontal surface in spacetime.
For $v \neq 0$, the lab's equal-time slice $t = 0$ transforms into
\be
t' = -\gamma v x, \qquad x' = \gamma x,
\ee
leading to $t'=-v x'$. This is no longer a constant-$t'$ surface, different spatial points $x$ in the lab correspond to different times $t'$ in the rest frame. Hence, in the rest frame, the boundary between the strip and its complement is not a purely spatial cut at one instant and it extends over a range of times. Because of the correlations between points at different times, this tilted cut intersects many more correlated pairs than a simple equal-time cut. The result is that EE can increase with velocity $v$ which may explain the increase observed in our model.

Regarding the effect of chemical potential, increasing $\mu$ adds more charged particles to the plasma. A higher density of charged degrees of freedom provides more quantum correlations that can cross the entangling surface between a subsystem and its complement. Consequently, the HEE may increase monotonically with $\mu$.

From the holographic perspective, for a strip oriented perpendicular to the boost direction, increasing either the velocity $v$ or the chemical potential $\mu$ does not necessarily push the turning point $z_*$ of the RT surface deeper into the bulk. However, since the final solution is described by the functions $x(z)$ and $t(z)$, as shown in figure \ref{general}, the bending of these curves becomes more pronounced. This increased curvature leads to a larger minimal surface area and since the HEE is proportional to this area the entanglement entropy increases with both $v$ and $\mu$.

A few comments are in order here:
\begin{itemize}
\item At fixed temperature, figure~\ref{figvmu0} shows that when either the velocity $v$ or the chemical potential $\mu$ is sufficiently small, the HEE does not change substantially even if the other parameter is varied. In fact, if one of them is small, the effect of the other becomes insignificant and the HEE remains  close to that of the pure thermal background.
\item At fixed temperature, for sufficiently large values of the chemical potential, the effect of high velocity is clearly observed. High enough velocity serves as a good criterion to distinguish different physical systems with various chemical potentials.
\item At high temperatures, the two effects work together: the temperature generates a large number of modes while the velocity enhances the correlations of those modes across the strip's boundary, resulting in a substantial increase.
\item Comparing the chemical potential and temperature, the temperature plays a more important role than the chemical potential; that is, the HEE increases more with $T$ than with $\mu$. This indicates that the system is more sensitive to thermal fluctuations variations than to charge density. Thus, the temperature almost becomes the primary control parameter for HEE.
\item In the ultrarelativistic regime, the HEE grows very rapidly with velocity which emerges as the dominant parameter and largely suppresses the effects of both temperature and chemical potential. This steep growth appears to reflect the kinematic dominance of the Lorentz factor $\gamma$ over thermodynamic contributions, making the HEE a highly sensitive probe of the boost in strongly coupled gauge theory.
\end{itemize}


\begin{thebibliography}{99}
\bibitem{Maldacena:1997re}
J.~M.~Maldacena,
``The Large $N$ limit of superconformal field theories and supergravity,''
Adv. Theor. Math. Phys. \textbf{2} (1998), 231-252
[arXiv:hep-th/9711200 [hep-th]].
\bibitem{Witten:1998qj}
E.~Witten,
``Anti de Sitter space and holography,''
Adv. Theor. Math. Phys. \textbf{2} (1998), 253-291
[arXiv:hep-th/9802150 [hep-th]].
\bibitem{Gubser:1998bc}
S.~S.~Gubser, I.~R.~Klebanov and A.~M.~Polyakov,
``Gauge theory correlators from noncritical string theory,''
Phys. Lett. B \textbf{428} (1998), 105-114
[arXiv:hep-th/9802109 [hep-th]].
\bibitem{Aharony:1999ti}
O.~Aharony, S.~S.~Gubser, J.~M.~Maldacena, H.~Ooguri and Y.~Oz,
``Large N field theories, string theory and gravity,''
Phys. Rept. \textbf{323} (2000), 183-386
[arXiv:hep-th/9905111 [hep-th]].
\bibitem{Headrick:2019eth}
M.~Headrick,
``Lectures on entanglement entropy in field theory and holography,''
[arXiv:1907.08126 [hep-th]].
\bibitem{Sorkin:1984kjy}
R.~D.~Sorkin,
``1983 paper on entanglement entropy: ''On the Entropy of the Vacuum outside a Horizon,''
[arXiv:1402.3589 [gr-qc]].
\bibitem{Calabrese:2004eu}
P.~Calabrese and J.~L.~Cardy,
``Entanglement entropy and quantum field theory,''
J. Stat. Mech. \textbf{0406} (2004), P06002
[arXiv:hep-th/0405152 [hep-th]].
\bibitem{Casini:2009sr}
H.~Casini and M.~Huerta,
``Entanglement entropy in free quantum field theory,''
J. Phys. A \textbf{42} (2009), 504007
[arXiv:0905.2562 [hep-th]].
\bibitem{Fischler:2012uv}
W.~Fischler, A.~Kundu and S.~Kundu,
``Holographic Mutual Information at Finite Temperature,''
Phys. Rev. D \textbf{87} (2013) no.12, 126012
[arXiv:1212.4764 [hep-th]].
\bibitem{Ali-Akbari:2013hba}
M.~Ali-Akbari and A.~Vahedi,
``Non-equilibrium Phase Transition from AdS/CFT,''
Nucl. Phys. B \textbf{877} (2013), 95-106
[arXiv:1305.3713 [hep-th]].
\bibitem{Witten:1998zw}
E.~Witten,
``Anti-de Sitter space, thermal phase transition, and confinement in gauge theories,''
Adv. Theor. Math. Phys. \textbf{2} (1998), 505-532
[arXiv:hep-th/9803131 [hep-th]].
\bibitem{Ahmed:2023dnh}
M.~B.~Ahmed, W.~Cong, D.~Kubiznak, R.~B.~Mann and M.~R.~Visser,
``Holographic CFT phase transitions and criticality for rotating AdS black holes,''
JHEP \textbf{08} (2023), 142
[arXiv:2305.03161 [hep-th]].
\bibitem{Matsumoto:2018ukk}
M.~Matsumoto and S.~Nakamura,
``Critical Exponents of Nonequilibrium Phase Transitions in AdS/CFT Correspondence,''
Phys. Rev. D \textbf{98} (2018) no.10, 106027
[arXiv:1804.10124 [hep-th]].
\bibitem{Vahedi:2018gvn}
A.~Vahedi and M.~Shakeri,
``Non-Equilibrium Critical Phenomena From Probe Brane Holography in Schr{\"o}dinger Spacetime,''
JHEP \textbf{01} (2019), 047
[arXiv:1811.05823 [hep-th]].
\bibitem{Lindgren:2017hiu}
E.~J.~Lindgren,
``Black hole formation, holographic thermalization and the AdS/CFT correspondence,''
[arXiv:1909.00434 [hep-th]].
\bibitem{Ryu:2006bv}
S.~Ryu and T.~Takayanagi,
``Holographic derivation of entanglement entropy from AdS/CFT,''
Phys. Rev. Lett. \textbf{96} (2006), 181602
[arXiv:hep-th/0603001 [hep-th]].
\bibitem{Headrick:2010zt}
M.~Headrick,
``Entanglement Renyi entropies in holographic theories,''
Phys. Rev. D \textbf{82} (2010), 126010
[arXiv:1006.0047 [hep-th]].
\bibitem{Tonni:2010pv}
E.~Tonni,
``Holographic entanglement entropy: near horizon geometry and disconnected regions,''
JHEP \textbf{05} (2011), 004
[arXiv:1011.0166 [hep-th]].
\bibitem{Fischler:2012uv}
W.~Fischler, A.~Kundu and S.~Kundu,
``Holographic Mutual Information at Finite Temperature,''
Phys. Rev. D \textbf{87} (2013) no.12, 126012
[arXiv:1212.4764 [hep-th]].
\bibitem{Molina-Vilaplana:2011ydi}
J.~Molina-Vilaplana and P.~Sodano,
``Holographic View on Quantum Correlations and Mutual Information between Disjoint Blocks of a Quantum Critical System,''
JHEP \textbf{10} (2011), 011
[arXiv:1108.1277 [quant-ph]].
\bibitem{Fischler:2013gsa}
W.~Fischler, A.~Kundu and S.~Kundu,
``Holographic Entanglement in a Noncommutative Gauge Theory,''
JHEP \textbf{01} (2014), 137
[arXiv:1307.2932 [hep-th]].
\bibitem{Ling:2015dma}
Y.~Ling, P.~Liu, C.~Niu, J.~P.~Wu and Z.~Y.~Xian,
``Holographic Entanglement Entropy Close to Quantum Phase Transitions,''
JHEP \textbf{04} (2016), 114
[arXiv:1502.03661 [hep-th]].
\bibitem{Kundu:2016dyk}
S.~Kundu and J.~F.~Pedraza,
``Aspects of Holographic Entanglement at Finite Temperature and Chemical Potential,''
JHEP \textbf{08} (2016), 177
[arXiv:1602.07353 [hep-th]].
\bibitem{Chamblin:1999tk}
A.~Chamblin, R.~Emparan, C.~V.~Johnson and R.~C.~Myers,
``Charged AdS black holes and catastrophic holography,''
Phys. Rev. D \textbf{60} (1999), 064018
[arXiv:hep-th/9902170 [hep-th]].
\bibitem{Cvetic:1999ne}
M.~Cvetic and S.~S.~Gubser,
``Phases of R charged black holes, spinning branes and strongly coupled gauge theories,''
JHEP \textbf{04} (1999), 024
[arXiv:hep-th/9902195 [hep-th]].
\bibitem{Cai:1996eg}
R.~G.~Cai and Y.~Z.~Zhang,
``Black plane solutions in four-dimensional space-times,''
Phys. Rev. D \textbf{54} (1996), 4891-4898
[arXiv:gr-qc/9609065 [gr-qc]].
\bibitem{Bhattacharya:2022msw}
J.~Bhattacharya, P.~Biswas, A.~Chandranathan and S.~K.~Das,
``Holographic entanglement entropy for relativistic hydrodynamic flows,''
JHEP \textbf{05} (2023), 092
[arXiv:2211.14271 [hep-th]].
\bibitem{Saha:2019ado}
A.~Saha, S.~Gangopadhyay and J.~P.~Saha,
``Holographic entanglement entropy and generalized entanglement temperature,''
Phys. Rev. D \textbf{100} (2019) no.10, 106008
[arXiv:1906.03159 [hep-th]].
\bibitem{Ebrahim:2023ush}
H.~Ebrahim and M.~Ahmadpour,
``Holographic entanglement entropy and mutual information in deformed field theories at finite temperature,''
Phys. Rev. D \textbf{107} (2023) no.8, 086010
[arXiv:2301.07242 [hep-th]].
\bibitem{Rahimi:2018ica}
M.~Rahimi and M.~Ali-Akbari,
``Holographic Entanglement Entropy Decomposition in an Anisotropic Gauge Theory,''
Phys. Rev. D \textbf{98} (2018) no.2, 026004
[arXiv:1803.01754 [hep-th]].
\bibitem{Bah:2007kcs}
I.~Bah, A.~Faraggi, L.~A.~Pando Zayas and C.~A.~Terrero-Escalante,
``Holographic entanglement entropy and phase transitions at finite temperature,''
Int. J. Mod. Phys. A \textbf{24} (2009), 2703-2728
[arXiv:0710.5483 [hep-th]].
\bibitem{Headrick:2007km}
M.~Headrick and T.~Takayanagi,
``A Holographic proof of the strong subadditivity of entanglement entropy,''
Phys. Rev. D \textbf{76} (2007), 106013
[arXiv:0704.3719 [hep-th]].
\bibitem{Headrick:2013zda}
M.~Headrick,
``General properties of holographic entanglement entropy,''
JHEP \textbf{03} (2014), 085
[arXiv:1312.6717 [hep-th]].










\end{thebibliography}
\end{document}